\documentclass[twocolumn]{aastex62}

\usepackage{amsmath}
\usepackage{comment}
\usepackage[normalem]{ulem}

\newcommand{\mmin}{m_\mathrm{min}}
\newcommand{\mmax}{m_\mathrm{max}}

\newcommand{\modelBalpha}{-1.6^{+1.7}_{-1.5}}
\newcommand{\modelBbeta}{6.7^{+4.8}_{-5.9}}
\newcommand{\modelBmmin}{7.9^{+1.2}_{-2.5} \ M_\odot}
\newcommand{\modelBmmax}{42.0^{+15.0}_{-5.7} \ M_\odot}

\newcommand{\LVCalpha}{-1.7^{+1.8}_{-1.5}}
\newcommand{\LVCbeta}{6.1^{+5.4}_{-5.7}}
\newcommand{\LVCmmin}{7.3^{+1.7}_{-3.7} \ M_\odot}
\newcommand{\LVCmmax}{41.8^{+18.6}_{-5.5} \ M_\odot}

\newcommand{\RPgamma}{-1.1^{+1.0}_{-0.9}}

\newcommand{\betabeta}{7.0^{+4.5}_{-5.5}}

\newcommand{\qmingamma}{-1.3^{+0.9}_{-0.8}}

\newcommand{\qminbetaMMax}{41.9^{+18.2}_{-5.7} \ M_\odot}

\newcommand{\qminbetaMMin}{6.7^{+1.9}_{-3.2} \ M_\odot}

\newcommand{\pbetagrzero}{0.987}

\newcommand{\lvcqonep}{0.59^{+0.13}_{-0.33}}
\newcommand{\RPqonep}{0.15^{+0.07}_{-0.06}}
\newcommand{\qminbetaqonep}{0.66^{+0.25}_{-0.28}}

\newcommand{\mocksixtyyscale}{0.5^{+0.4}_{-0.4}}
\newcommand{\mocksixtygamma}{-1.1^{+0.4}_{-0.3}}
\newcommand{\mocksixtybeta}{3.4^{+4.6}_{-5.9}}
\newcommand{\mocksixtyMMin}{6.5^{+1.3}_{-2.4}}
\newcommand{\mocksixtyMMax}{40.6^{+2.4}_{-1.7}}
\newcommand{\mocksixtyqmin}{0.5^{+0.3}_{-0.3}}

\newcommand{\mockhundredMMax}{38.9^{+1.4}_{-0.9} \ M_\odot}

\newcommand{\mockhundredgamma}{-0.8^{+0.3}_{-0.3}}

\newcommand{\pp}[1]{\textcolor{black}{#1}}

\graphicspath{{./}{figures/}}

\begin{document}

\title{Picky Partners: The Pairing of Component Masses in Binary Black Hole Mergers}

\author{Maya Fishbach}
\affiliation{{Department of Astronomy and Astrophysics, University of Chicago, Chicago, IL 60637, USA}}
\email{mfishbach@uchicago.edu}

\author{Daniel E. Holz}
\affiliation{Enrico Fermi Institute, Department of Physics, Department of Astronomy and Astrophysics,\\and Kavli Institute for Cosmological Physics, University of Chicago, Chicago, IL 60637, USA}

\begin{abstract}
We examine the relationship between individual black hole (BH) masses in merging binary black hole (BBH) systems. Analyzing the ten BBH detections from LIGO/Virgo's first two observing runs, we find that the masses of the component BHs comprising each binary are unlikely to be randomly drawn from the same underlying distribution. Instead, the two BHs of a given binary prefer to be of comparable mass. We show that it is $\sim 5$ times more likely that the component BHs in a given binary are always equal (to within 5\%) than that they are randomly paired. If we assume that the probability of a merger between two BHs scales with the mass ratio $q$ as $q^\beta$, so that $\beta=0$ corresponds to random pairings, we find $\beta>0$ is favored at credibility $\pbetagrzero$. 
By modeling the mass distribution, we find that the median mass ratio is $q_{50\%} = 0.91^{+0.05}_{-0.17}$ at 90\% credibility.
\added{While the pairing between BHs depends on their mass ratio, we find no evidence that it depends on the total mass of the system: it is $\sim 6$ times more likely that the pairing depends purely on the mass ratio than on the total mass.} We predict that 99\% of BBHs detected by LIGO/Virgo will have mass ratios $q > 0.5$.
We conclude that merging black holes do not form random pairings; instead they are selective about their partners, preferring to mate with black holes of a similar mass. The details of these selective pairings provide insight into the underlying formation channels of merging binaries.
\end{abstract}

\section{Introduction} \label{sec:intro}
Following the first two observing runs (O1 and O2) of advanced LIGO~\citep{TheLIGOScientific:2014jea} and Virgo~\citep{TheVirgo:2014hva}, the LIGO/Virgo Collaboration (LVC) reported ten detections of merging binary black holes (BBH)~\citep{Abbott:catalog}, with tens more detections expected from the third observing run (O3), and hundreds of expected detections per year once the gravitational-wave (GW) detector network reaches design sensitivity~\citep{ObsScen}.
\added{In addition to the LVC-published detections of~\cite{Abbott:catalog}, new BBH detections in the O1 and O2 data have been reported by \cite{IAS:O1}, \cite{IAS:O2}, and \cite{Nitz:Catalog}. }
The formation and history of these BBHs remains a fundamental question in GW astrophysics.
The proposed formation channels include isolated~\citep{2015ApJ...806..263D, 2016Natur.534..512B,  2016ApJ...824L..10W, 2016MNRAS.462.3302E, 2016ApJ...819..108B, Stevenson:2017, 2018MNRAS.481.1908K, 2019MNRAS.485..889S}, dynamical~\citep{2016MNRAS.459.3432M, 2016PASA...33...36H, 2016ApJ...824L...8R, 2017MNRAS.464L..36A, 2017ApJ...836L..26C, 2018PhRvL.120o1101R, 2018PhRvD..97j3014S, 2019ApJ...871...91Z, 2019arXiv190100863D}, and primordial~\citep{2016PhRvL.116t1301B,2017JPhCS.840a2032G}, with many variants within each model.
Different formation channels are expected to leave an imprint on the properties of the BBH population~\citep{2018MNRAS.477.4685B,2018PhRvD..98h3017T,2019MNRAS.482.2991A}, including the mass distribution~\citep{Stevenson:2015, Zevin:2017}, spin distribution~\citep{Rodriguez:spins,Farr:2017,Vitale:spins,2018ApJ...854L...9F}, and redshift evolution~\citep{Redshift,Vitale:2018,2018ApJ...866L...5R}. It is therefore possible to learn about the astrophysics of BBH formation by fitting for these population distributions using GW data. In~\cite{Abbott:pop}, the LVC carried out such an analysis on the first ten BBH detections, fitting the mass, spin and redshift distributions with simple parameterized models. For example, the mass distribution was fit to a model in which the primary mass (the more massive component of a binary) follows a power-law between some minimum and maximum mass, while the secondary mass is distributed with a power-law between the minimum mass and its primary mass partner. \cite{Abbott:pop} additionally considered a slightly more complex model, which replaces the minimum mass cutoff with tapering at the low-mass end and allows for an additional Gaussian component at the high mass end of the primary-mass power-law. In this work, we restrict the population analysis to the ten~\cite{Abbott:catalog} BBH detections, as the detection efficiency has been previously studied for this sample and is well-understood. \added{The detection efficiency between the LVC detections and the \cite{IAS:O2} detections is significantly different; see e.g. Figure 5 of \cite{IAS:O2}.} Using the wrong detection efficiency leads to selection biases in population inference. In future work we will extend our analysis to include overlapping samples with differing selection effects.

In this work we extend the analysis of~\cite{Abbott:pop} by focusing on a particular aspect of the BBH mass distribution: the pairing between the two component BHs in the binary. 
We ask whether the universe makes merging binary black hole systems by randomly pairing up black holes, or whether the mass of each black hole in a pair influences the mass of its companion. 
This differs from the analysis of~\cite{Abbott:pop}, in which the parameterization for the mass distribution does not separate the underlying BH mass distribution and the pairing function. Under the models considered by~\cite{Abbott:pop}, it is not possible to fit for an underlying mass distribution that is common to both component BHs or quantify the deviation from the random-pairing scenario, as we do in this work.

We expect that the pairing function carries an imprint of the physics by which component BHs find their partners. \added{Various formation models predict that the mass ratio and/or total mass of the two components may determine their probability of merging.} Despite the different physical processes involved, many formation channels predict a preference for similar component masses~\citep{2018arXiv180605820M}.
Binaries formed via homogeneous chemical evolution are expected to strongly prefer equal mass components \added{due to the progenitor stars exchanging mass during an early overcontact phase~\citep{2016MNRAS.458.2634M,2016A&A...588A..50M}}.
The traditional isolated evolution channel is also expected to favor comparable mass components, \added{because the common envelope phase is unsuccessful at producing close binaries for extreme mass ratio systems~\citep{2015ApJ...806..263D}}. However, the common envelope phase remains poorly understood, and this channel can produce mergers between fairly unequal component masses, especially at lower metallicities~\citep{Dominik:2012,Stevenson:2017,Klencki:2018,Giacobbo:2018,Spera:2019}. Some studies have suggested that dynamical evolution also tends to produce more mergers with equal mass components due to the fact that comparable mass binaries have a higher binding energy and form tighter binaries~\citep{Rodriguez:2016a,2016MNRAS.458.3075A}. However, other dynamical channels may mildly prefer unequal mass components~\citep{2019arXiv190201864M}. Alternatively, it has been suggested that in dynamical channels, the merger probability depends on the total mass, rather than the mass ratio, as mass segregation and dynamical interactions may favor binaries with larger total masses~\citep{2016ApJ...824L..12O,2019arXiv190103345P}. \added{\cite{2018ApJ...854...41K} proposed measuring the pairing function's dependence on the total mass to discriminate between formation channels, as the pairing function is expected to scale as $M_\mathrm{tot}^\alpha$ with $\alpha = 4$ in the dynamical channel modeled by~\cite{2016ApJ...824L..12O} and $\alpha \sim 1$ for merging primordial BHs.} Constraining the BBH pairing function with GW observations allows us to test these different predictions.

The pairing function has been previously studied in the context of the initial mass function for binary stars, where the degree of correlation between component stars (and the dependence on the orbital separation) remains an open question~\citep{2006ApJ...639L..67P,2009A&A...493..979K,2013pss5.book..115K,2017ApJS..230...15M,2018arXiv180610605K}. It is possible that studying the pairing function for merging BBHs may shed light on the masses of their stellar progenitors, although the relationship between a BH's mass and its progenitor star's zero-age main-sequence (ZAMS) mass is complicated by the many stages of evolution undergone by BBHs.

In the stellar context, it has been pointed out that different pairing functions affect the 1-dimensional distribution of mass ratios as well as the 1-dimensional distributions of primary and secondary masses~\citep{2009A&A...493..979K}. Because the primary (secondary) mass is defined to be the more (less) massive component in the binary, even randomly drawing two components from the same underlying distribution results in the primary and secondary masses having different distributions. Random draws can also result in very different mass ratio distributions, depending on the shape of the underlying mass distribution. We emphasize that the pairing mechanism cannot be determined by examining any one of these one-dimensional distributions independently. For example, a mass ratio distribution that favors near-unity mass ratios may simply indicate that the underlying BH mass distribution peaks in a narrow mass range, rather than that similar component masses are more likely to partner and merge. It is therefore important to examine the two-dimensional mass distribution in order to analyze whether or not there is a preference for similar-mass components. 

This paper explores the BBH pairing function by analyzing the first ten LIGO/Virgo BBHs according to the mass models described in Section~\ref{sec:models}. The results of the analysis and implications for future detections are found in Section~\ref{sec:results}. In Section~\ref{sec:simulations} we demonstrate the analysis on mock GW data and forecast the constraints that will be possible with $\sim 50$--$100$ more BBH detections (to be expected at the end of O3 or shortly after the start of O4). We conclude in Section~\ref{sec:conclusion}. Appendix~\ref{sec:methods} describes the details of the hierarchical Bayesian analysis.

\section{Mass Distribution Models} \label{sec:models}
In the simplest case, we consider a model in which the component masses in a BBH system are independently drawn from the same underlying power-law distribution:
\begin{equation} \label{eq:m}
    p(m) \propto m^\gamma, \ \ \mmin < m < \mmax,
\end{equation}
where $\gamma$ is the power-law slope, and $\mmin$ and $\mmax$ are the minimum and maximum mass.
We refer to this as the ``random pairing" mass distribution~\citep{2009A&A...493..979K}.
We note that in this case, the marginal distributions of the primary and secondary masses are not identical, because the primary (secondary) is defined as the more massive (less massive) component.
Defining $m_1$ as the primary mass and $m_2$ as the secondary mass, the random pairing power-law distribution takes the form:
\begin{widetext}
\begin{equation} \label{eq:RP}
    p(m_1, m_2 \mid \gamma, \mmin, \mmax) =
\begin{cases}
\frac{2  \left( \gamma+1 \right) ^2}{\left( \mmax^{\gamma+1}-\mmin^{\gamma+1}\right) ^2} m_1^\gamma m_2^\gamma  &\mbox{if } \mmin < m_2 < m_1 < \mmax \\
0 &\mbox{else}.
\end{cases}
\end{equation}
\end{widetext}
This implies that the primary masses follow the distribution:
\begin{equation}  \label{eq:m1}
    p(m_1 \mid \gamma, \mmin, \mmax) = \frac{2 \left(\gamma+1\right) m_1^\gamma \left(m_1^{\gamma+1}-\mmin^{\gamma+1}\right)}{\left(\mmax^{\gamma+1}-\mmin^{\gamma+1} \right)^2},
\end{equation}
while the secondaries follow:
\begin{equation} \label{eq:m2}
    p(m_2 \mid \gamma, \mmin, \mmax) = \frac{ 2\left(\gamma+1\right) m_2^\gamma \left(\mmax^{\gamma+1}-m_2^{\gamma+1}\right) }{\left( \mmax^{\gamma+1}-\mmin^{\gamma+1} \right)^2}.
\end{equation}
We reiterate that the distributions in Eqs.~\ref{eq:m1} and~\ref{eq:m2} are not the same as the underlying distribution (Eq.~\ref{eq:m}), even though both masses are separately drawn from this distribution. In particular, the primary mass distribution will tend to favor larger masses compared to the secondary mass distribution.
Furthermore, different choices of the underlying power-law parameters ($\gamma$, $\mmin$, and $\mmax$) will lead to different distributions in the mass ratio $q \equiv m_2 / m_1 \leq 1$. If the underlying power-law is steep enough in either direction, mass ratios close to unity will be favored even if the two components are randomly paired.

In order to explore the pairing of two component BHs, we consider mass distributions that contain the random pairing distribution as a sub-model, but allow for deviations parameterized by a pairing function, $f_p$. \added{Motivated by population synthesis models, we consider two pairing functions: one that depends on the mass ratio, $q$, where $q = m_2/ m_1 \leq 1$, and one that depends on the total mass, $M_\mathrm{tot} = m_1 + m_2$. We also consider the possibility that the probability of two BHs forming a binary and merging depends on both $q$ and $M_\mathrm{tot}$.}

If each BH mass is drawn from an underlying power-law distribution with slope $\gamma$, and the probability of two masses belonging to a merging binary is given by $f_p(q, M_\mathrm{tot} \mid \vec{\beta})$, where $\vec{\beta}$ denotes the hyper-parameter(s) of the pairing function, the mass distribution of merging BHs follows:
\begin{widetext}
\begin{equation}
    p(m_1, m_2 \mid \gamma, \mmin, \mmax, \vec{\beta}) \propto \begin{cases} m_1^\gamma m_2^\gamma f_p(\frac{m_2}{m_1}, m_1 + m_2 \mid \vec{\beta})&\mbox{if } \mmin < m_2 < m_1 < \mmax \\
0 &\mbox{else}.
    \end{cases}
\end{equation}
\end{widetext}

We first consider pairing functions, $f_p$, that depend purely on the mass ratio. As a simple model, we assume that the pairing function follows a power-law in mass ratio with slope $\beta_q$, with a minimum mass ratio threshold required for merger, $q_\mathrm{min}$. \added{The parameter $q_\mathrm{min}$ allows us to explore the scenario in which mergers only take place between equal component masses ($q_\mathrm{min} \rightarrow 1$).} In this model:
\begin{equation}
\label{eq:betaq}
f_p(q, M_\mathrm{tot} \mid \vec{\beta} = \lbrace \beta_q, q_\mathrm{min} \rbrace) \propto 
\begin{cases}
q^{\beta_q} &\mbox{if } q > q_\mathrm{min} \\
0 &\mbox{else.}
\end{cases}
\end{equation}
This model reduces to the random pairing model for $\left( \beta_q, q_\mathrm{min} \right) = \left( 0, \frac{\mmin}{\mmax} \right)$.

Second, we consider pairing functions $f_p$ that depend on the total mass of the system. A simple model in this case, motivated by the predictions of the dynamical channel of~\cite{2016ApJ...824L..12O} and the primordial BH channel of~\cite{2018ApJ...854...41K}, is a simple power-law in $M_\mathrm{tot}$:
\begin{equation}
\label{eq:betam}
    f_p(q, M_\mathrm{tot} \mid \vec{\beta} = \lbrace \beta_M \rbrace) \propto M_\mathrm{tot}^{\beta_M}.
\end{equation}
More generally, we may consider a pairing function that depends on both the mass ratio and the total mass:
\begin{equation}
\label{eq:betaqbetam}
f_p(q, M_\mathrm{tot} \mid \vec{\beta} = \lbrace \beta_q, q_\mathrm{min}, \beta_M \rbrace) \propto 
\begin{cases}
q^{\beta_q} M_\mathrm{tot}^{\beta_M} &\mbox{if } q > q_\mathrm{min} \\
0 &\mbox{else.}
\end{cases}
\end{equation}

We highlight that because we consider models that reduce to random pairing under some choice of parameters, our assumed parameterization differs from the mass distribution models analyzed in~\citet{Abbott:pop}. The basic power-law model in~\citet{Abbott:pop} is defined such that the marginal $p(m_1)$ distribution follows a power-law, so that the joint mass distribution takes the form~\citep[note we have redefined $\alpha$ from][as $-\alpha$]{Abbott:pop}:
\begin{equation} \label{eq:massLVC}
\begin{split}
    p(m_1,m_2 &\mid \alpha, \beta, \mmin \mmax) = \\ &(\alpha+1)(\beta+1)\frac{m_1^{\alpha}}{\mmax^{\alpha+1}-\mmin^{\alpha+1}}\frac{m_2^{\beta}}{m_1^{\beta+1}-\mmin^{\beta+1}}.
\end{split}
\end{equation}
On the other hand, for the parameterizations we consider in this work, the marginal distribution of primary masses does not follow an exact power-law; instead, these parameterizations allow for the possibility that both masses in a binary are drawn from the same underlying power-law distribution.

Following the methods laid out in Appendix~\ref{sec:methods}, we fit the pairing models discussed above to the first ten BBH detections in Section~\ref{sec:results}. \added{As part of our results, we quantify the evidence against the random pairing and the total-mass dependent pairing models, and find that the mass-ratio dependent pairing model provides the best fit to the data.}

\section{Results}
\label{sec:results}
\subsection{LVC Model}
We begin by recovering the results of~\cite{Abbott:pop} under the same mass model given by Eq.~\ref{eq:massLVC}, equivalent to Model B of~\cite{Abbott:pop}, in order to demonstrate consistency between our methods. Although we use slightly different assumptions regarding the spin distribution and the selection effects calculation (see Appendix~\ref{sec:methods}), we recover nearly-identical posterior distributions on the population hyper-parameters: $\alpha = \LVCalpha$, $\beta = \LVCbeta$, $\mmin = \LVCmmin$, and $\mmax = \LVCmmax$. This is to be compared with $\alpha = \modelBalpha$, $\beta = \modelBbeta$, $\mmin = \modelBmmin$, and $\mmax = \modelBmmax$ found by~\citet{Abbott:pop}. As described in Appendix~\ref{sec:methods}, our prior on $\mmin$ starts at 3 $M_\odot$ rather than 5 $M_\odot$. Furthermore, recall that our convention for the power-law slope $\alpha$ has a sign flip compared to the convention in~\citet{Abbott:pop}. 

With the current set of events, the data cannot distinguish between the mass model of~\cite{Abbott:pop} (Eq.~\ref{eq:massLVC}) and the models we consider in this work, and they all give consistent results for the inferred mass distribution $p(m_1, m_2)$. However, the parameters of our models have a different interpretation from the mass model of~\cite{Abbott:pop}. While the power-law slope $\alpha$ of Eq.~\ref{eq:massLVC} refers to the power-law of the primary mass distribution, the power-law slope $\gamma$ of this work refers to the underlying mass distribution power-law from which both primary and secondary BHs are drawn. The additional parameters $\beta_q$, $q_\mathrm{min}$ and $\beta_M$ in our models allow us to explore whether the pairing between the two component masses is random or whether (and how) it depends on the mass ratio \added{or total mass of the system}, according to Eq.~\ref{eq:betaqbetam}). 

\begin{figure*}
\includegraphics[width=\textwidth]{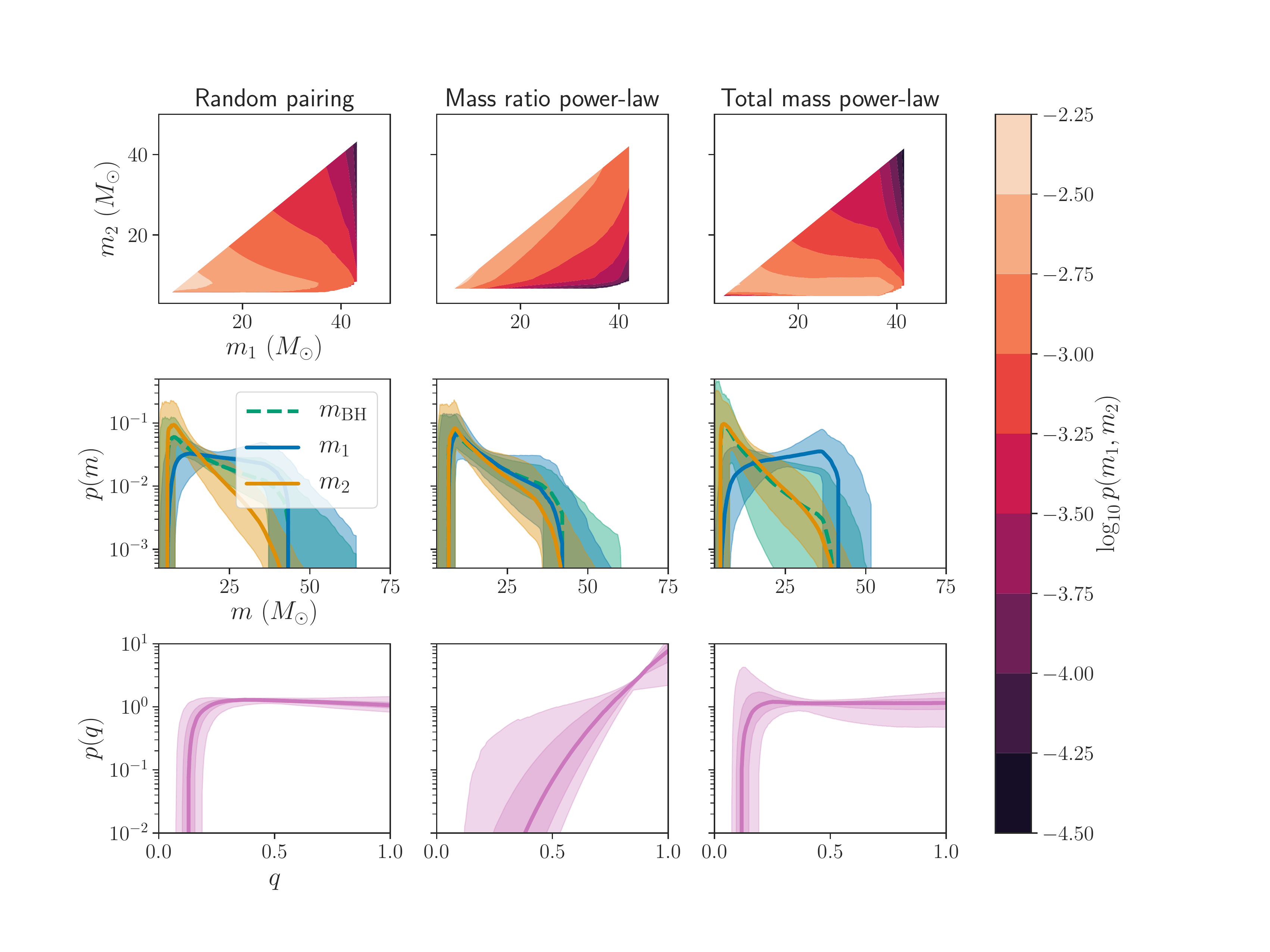}
\caption{\label{fig:pm1m2} \emph{Top row:} Joint $m_1$--$m_2$ distribution as inferred from the ten BBHs assuming a mass distribution given by Eq.~\ref{eq:betaqbetam} with free parameters $\gamma$, $\mmin$, $\mmax$ (left column), $\gamma$, $\mmin$, $\mmax$ and $\beta_q$ (middle column), \added{and $\gamma$, $\mmin$, $\mmax$ and $\beta_M$ (right column)}. In each case, those parameters that are not free are fixed to $\beta_q = \beta_M = 0$ and $q_\mathrm{min} = \mmin / \mmax$. The color scale indicates the median $\log_{10}$ of the merger rate density as a function of the two masses. \emph{Middle row:} Marginal distributions of single BH masses (green), along with the primary masses (blue) and secondary masses (yellow) of component BHs in binary systems. These distributions are inferred by fitting the ten BBH detections to the model of the corresponding column. The line shows the median merger rate density as a function of mass, while the shaded bands show symmetric 90\% credible intervals. \emph{Bottom row:} Marginal distribution of the mass ratio implied by the fits to the three models. The solid line and dark (light) bands denote median and 50\% (90\%) credible intervals on the merger rate as a function of mass ratio.}
\end{figure*}

\subsection{Random Pairing}
\label{sec:methodsRP}
Figure~\ref{fig:pm1m2} shows the results of fitting the random-pairing model (Eq.~\ref{eq:RP} with free parameters $\gamma$, $\mmin$, $\mmax$), the mass ratio power-law model (Eq.~\ref{eq:betaq} with free parameters $\gamma$, $\mmin$, $\mmax$, $\beta_q$) and the total mass power-law (Eq.~\ref{eq:betam} with free parameters $\gamma$, $\mmin$, $\mmax$, $\beta_M$) to the ten BBHs from the first two observing runs. We use flat priors on all free parameters, with uninformative prior bounds listed in Appendix~\ref{sec:methods}. In each case, the parameters of Eq.~\ref{eq:betaqbetam} that are not left free are fixed to the default values $\beta_q = \beta_M = 0$, $q_\mathrm{min} = \mmin / \mmax$.

If we fix the pairing to be random, we find $\gamma = \RPgamma$. However, as shown in the left-hand, middle panel in Figure~\ref{fig:pm1m2}, and explained in Section~\ref{sec:models}, this does not imply that the one-dimensional marginal distributions of the primary and secondary masses follow this common power-law; the primary masses follow a flatter distribution while the secondary masses follow a steeper distribution. Note that the inferred mass ratio distribution (bottom row, left-hand panel) in this case is inferred to be nearly flat across the range $\sim 0.15$--$1$. This is a consequence of this particular fit to the random-pairing model; in general the marginal mass ratio distribution can slope significantly upwards depending on the value of $\gamma$. The lower-limit on the mass ratio in the random-pairing model is given by the ratio $\mmin/\mmax$, \pp{and is constrained to $\sim 0.15$ in this case due to the measurements $\mmin \sim 7 \ M_\odot$ and $\mmax \sim 40 \ M_\odot$}.

\subsection{Mass Ratio Dependent Pairing}
\label{sec:resultsq}
The effect of introducing $\beta_q$ as a free parameter, while still fixing $q_\mathrm{min} = \mmin / \mmax$ and $\beta_M = 0$, is shown the middle column of Figure~\ref{fig:pm1m2}. Under this model extension, the data display a clear preference for mass ratios close to unity (bottom row, middle column), which implies more overlap between the primary and secondary mass distributions (middle row, central panel). We infer $\beta_q = \betabeta$, and find that $\beta_q \leq 0$ is ruled out with probability $\pbetagrzero$. This suggests that \emph{the random pairing model ($\beta_q = 0$) is strongly disfavored by the data.}
Meanwhile, the underlying mass distribution power-law slope is inferred to be a bit steeper than in the random-pairing case, with $\gamma = \qmingamma$ compared to $\gamma = \RPgamma$ for the random-pairing model. All models essentially agree on $\mmin \sim \qminbetaMMin$ and $\mmax \sim \qminbetaMMax$. 

\begin{figure*}
\includegraphics[width=\textwidth]{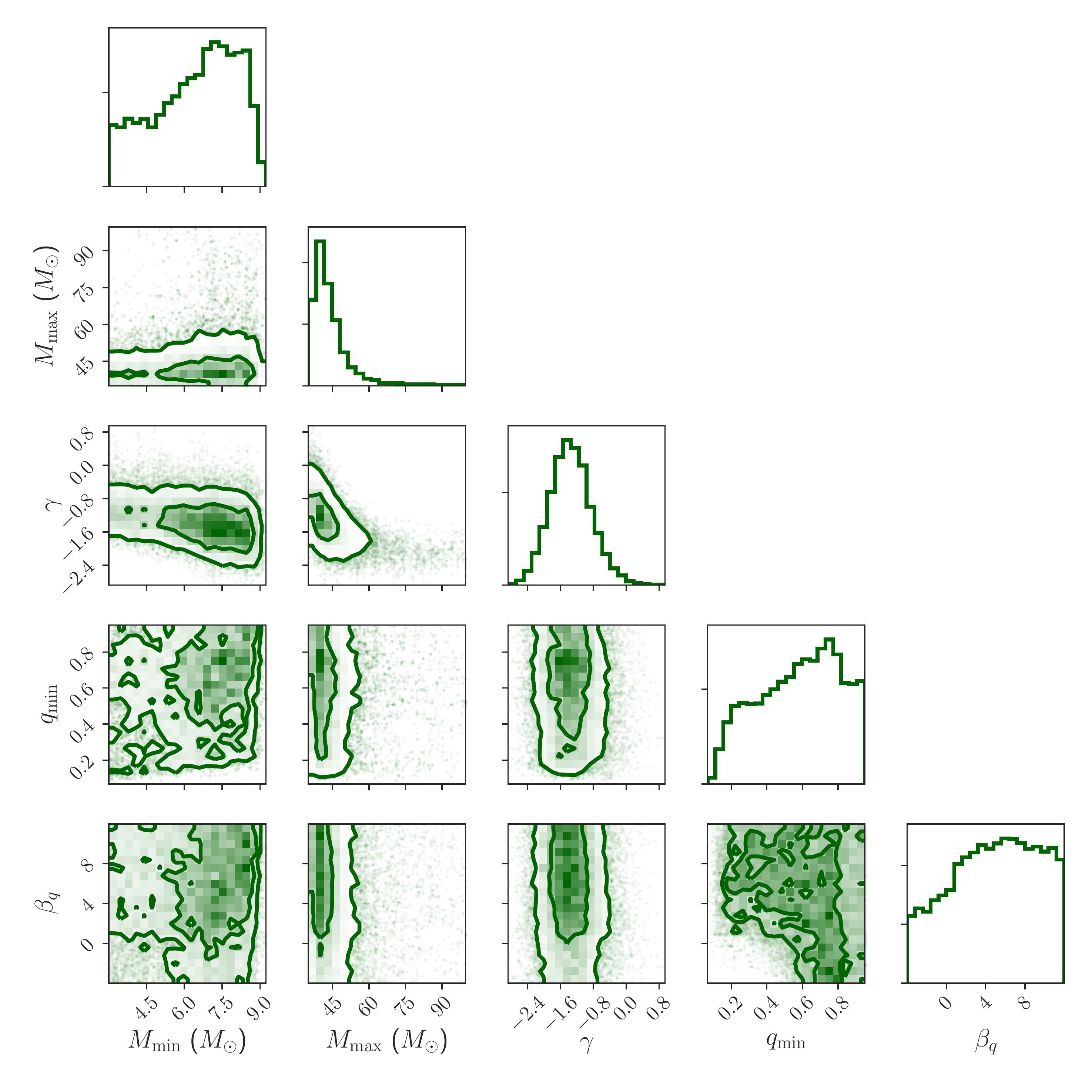}
\caption{\label{fig:corner_qminbeta} Posterior on the hyper-parameters of the power-law model with the mass-ratio dependent pairing (Eq.~\ref{eq:betaq}) fit to the ten BBH detections from O1 and O2. In the two-dimensional plots, the contours denote 50\% and 90\% posterior credible regions.}
\end{figure*}

We fit the model with both $\beta_q$ and $q_\mathrm{min}$ left free in Figure~\ref{fig:corner_qminbeta}, which displays the posterior distributions on the five hyper-parameters (three parameters to characterize the 1-dimensional mass distribution, and two -- $\beta_q$ and $q_\mathrm{min}$ -- to characterize the pairing function) as a corner plot~\citep{corner}. We find a strong preference for near-equal mass ratios, inferring that 99\% of merging BBHs have mass ratios between $q_{1\%} = \qminbetaqonep$ and unity. For reference, under Model B from~\cite{Abbott:pop}, we find $q_{1\%} = \lvcqonep$. Meanwhile, we find that the median mass ratio is $q_{50\%} = 0.91^{+0.05}_{-0.17}$. This agrees very closely with the findings of~\cite{Roulet:2019}, who find a preference for population distributions with an average mass ratio $\bar{q} =  0.89^{+0.08}_{-0.18}$. 
In fact, {the current set of detections is consistent with all binaries consisting of equal component masses} (to within 5\%, with $q_\mathrm{min} = 0.95$; the maximum value permitted by our prior). We find that $q_\mathrm{min} = 0.95$ is five times more likely than $(\beta_q, q_\mathrm{min}) = (0, \mmin/\mmax)$, meaning that it is {\em five times more likely that all binaries consist of equal component masses than that they are randomly paired}.
Although we do not include the events of~\citet{IAS:O1,IAS:O2} or~\citet{Nitz:Catalog} in our analysis (\added{in order to avoid assuming an incorrect selection function and biasing our results}), we note that all of their detections are also consistent with mass ratios of unity. This suggests an even stronger preference for near-equal component masses in the underlying population. 

\added{A strong preference for near-equal component masses with $q_\mathrm{min} \gtrsim 0.9$ is consistent with the chemically homogeneous/massive overcontact binary evolutionary channel~\citep{2016A&A...588A..50M}. However, our results are also consistent with a milder preference for near-unity mass ratios, which may be expected from classical isolated binary evolution or dynamical formation~\citep{2018arXiv180605820M}.}

\added{\subsection{Total Mass Dependent Pairing}
The effect of introducing $\beta_M$ as a free parameter, while fixing $q_\mathrm{min} = \mmin / \mmax$ and $\beta_q = 0$, is shown in the right-hand column of Figure~\ref{fig:pm1m2}. We note that these results are similar to the random pairing case, as there is a strong degeneracy between the isolated BH mass power-law slope $\gamma$ and the pairing function power-law slope $\beta_M$. We do not recover significant constraints beyond the linear combination $2\gamma + \beta_M \approx \RPgamma$ (matching the constraints reported in Section~\ref{sec:methodsRP} for the case $\beta_M = 0$). The degeneracy between $\beta_M$ and $\gamma$ causes the constraints on the isolated BH mass function (shown in green in the left-hand column, middle row of Figure~\ref{fig:pm1m2}) to be degraded. In addition to the degraded constraints on $\gamma$, the constraints on $\mmin$ are also less informative, as there is a significant correlation between $\mmin$ and $\gamma$, with shallow (less negative) power-law slopes corresponding to smaller allowed values of $\mmin$. However, $\mmax$ remains very well-measured at $\mmax \sim 41.9$.

It is interesting to note the inferred value of the power-law slope $\gamma$ for the cases $\beta_M = 4$ (predicted by some globular cluster simulations) and $\beta_M = 1$ (predicted by primordial BH channel). For $\beta_M = 4$, we find $\gamma \sim -2.5$, which is close to the Salpeter initial mass function for massive stars. The interpretation in this case is that BHs in globular clusters, where the merger probability roughly follows $M_\mathrm{tot}^4$~\citep{2006ApJ...637..937O}, are distributed according to a power-law mass distribution with slope $\gamma \sim -2.5$. This agrees with the findings of~\cite{2019arXiv190103345P}, who forward-model dynamical mergers for various initial BH mass functions and compare the model predictions to the LIGO/Virgo data to infer an initial BH mass function with slope $-2.35^{+0.36}_{-0.55}$ (68\% credibility). Meanwhile, $\beta_M = 1$ implies a shallower underlying mass distribution for single BHs, with $\gamma \sim -1$.

If we leave all three parameters of the pairing function, $q_\mathrm{min}$, $\beta_q$, and $\beta_M$, free, we recover identical constraints on $q_\mathrm{min}$ and $\beta_q$ as we do in the $\beta_M = 0$ case. The additional freedom in $\beta_M$ is fully absorbed by the degeneracy with $\gamma$, and we recover the prior on $\beta_M$. This is expected, because the data prefers equal mass systems with $m_1 = m_2$, and in this limit, Eq.~\ref{eq:betaqbetam} reduces to:
\begin{equation}
    p(m_1, m_2) \propto m_1^{2\gamma-\beta_M}, \mbox{ for } m_1 = m_2
\end{equation}
yielding a complete degeneracy along $2\gamma - \beta_M = \mathrm{const}$.}

\added{\subsection{Comparison of Pairing Functions}
The data strongly prefers a mass-ratio dependent pairing function that favors mergers between similar-mass components over random pairing, as discussed in Section~\ref{sec:resultsq}. However, unless one has a strong prior on the single BH mass distribution that would push $\beta_M$ away from zero (e.g. a prior belief that $\gamma < -2$ favors $\beta_M > 0$), there is no evidence that the total mass plays a role in the pairing function. This is apparent by the fact that when we fit Eq.~\ref{eq:betaqbetam} to the data with all six hyper-parameters free, we recover the prior on $\beta_M$.
Computing the evidence ratio in favor of a model in which $\beta_M = 0$ to one in which $(\beta_q, q_\mathrm{min}) = (0, \mmin / \mmax)$, we find that it is $\sim 6$ times more likely that the pairing function depends on the mass ratio than on the total mass.}

\begin{figure}
    \centering
    \includegraphics[width=0.5\textwidth]{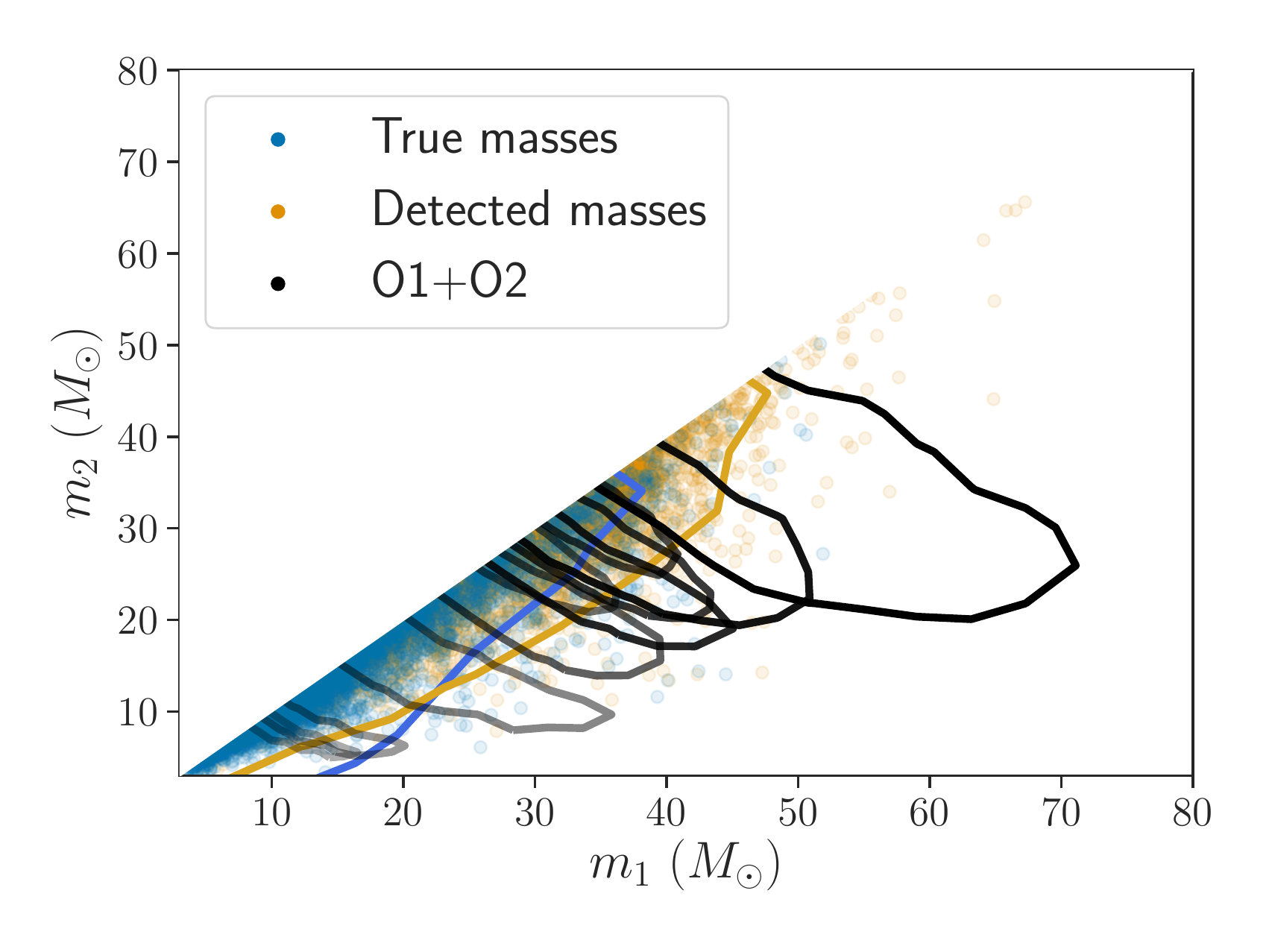}
    \caption{Posterior population distribution of the component masses in BBH binaries, as inferred from the mass-ratio dependent pairing model. The true masses of the \emph{underlying} population are represented by the blue points and 90\% credible region, while the orange points represent the \emph{detected} population, accounting for selection effects that favor more massive systems. In grayscale are the mass measurements of the ten LIGO/Virgo O1 and O2 detections. The contours denote 90\% credible intervals. All detected systems are consistent with equal component masses $m_1 = m_2$.}
    \label{fig:m1m2PPD}
\end{figure}

\subsection{Posterior predictive distributions}
In the following, because there is no evidence that the data prefers nonzero $\beta_M$, we fix $\beta_M = 0$ in the population model of Eq.~\ref{eq:betaqbetam} (reducing to the model of Eq.~\ref{eq:betaq}).
Using the recovered posteriors on the population hyper-parameters, shown as a corner plot in Figure~\ref{fig:corner_qminbeta}, we calculate the posterior population distribution $p(m_{1}, m_{2} \mid \rm data)$, shown in Figure~\ref{fig:m1m2PPD}. We define the posterior population distribution as in~\cite{Abbott:pop}; this refers to the distribution of true mass values marginalized over the hyper-parameter posteriors for a given population model:
\begin{equation}
    p(m_1, m_2 \mid d) = \int p(m_1, m_2 \mid \theta)p(\theta \mid d) \mathrm{d} \theta.
\end{equation}
Here, $p(\theta \mid d)$ refers to the posterior distribution on the population model's hyper-parameters inferred from the ten BBH events. Figure~\ref{fig:m1m2PPD} shows draws from the posterior population distribution on the true masses of the underlying BBH population (blue) as well as the true masses of {\em detected} systems (orange), found by applying selection effects to the underlying mass distribution. For comparison, the mass measurements of the O1 and O2 detections are shown in black.

\begin{figure}
\includegraphics[width=0.5\textwidth]{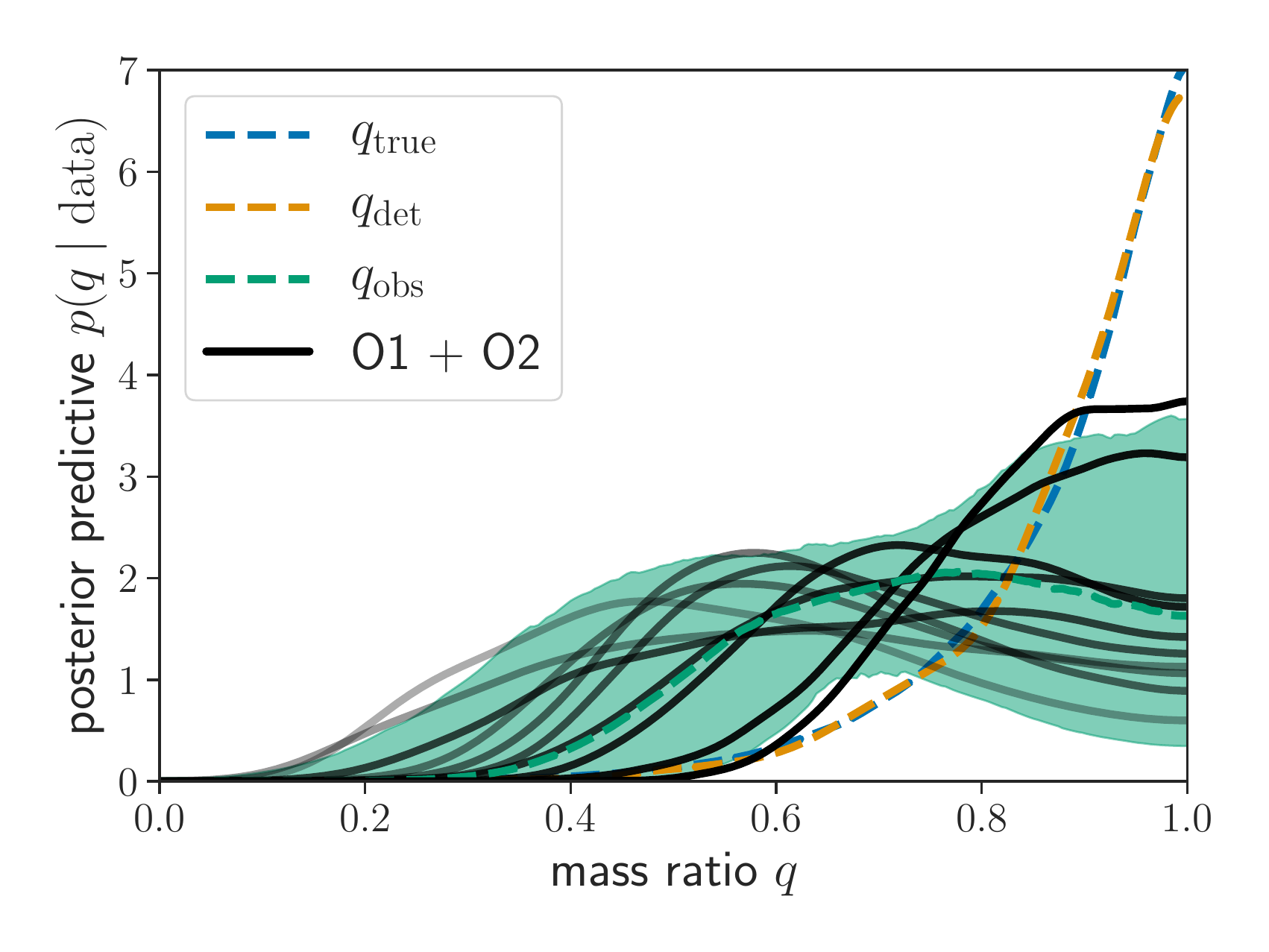}
\caption{\label{fig:PPD} Posterior population distribution of the mass ratio $q_\mathrm{true}$ in the \emph{underlying} population (dashed blue line), the mass ratio $q_\mathrm{det}$ among \emph{detected} systems (dashed orange line), and the posterior predictive process of the \emph{measured} mass ratio $q_\mathrm{obs}$ (dashed green line and shaded band), accounting for detection efficiency and measurement uncertainty. These distributions are inferred by fitting the ten BBHs from O1 and O2 to the mass distribution model described by Eq.~\ref{eq:betaq}. If all BBHs belong to this population, we expect that 90\% of the recovered posteriors from detected BBHs will fall within the shaded green region. The grayscale lines show the posterior probability distributions of the ten observed BBHs. Note that measurement uncertainty shifts the posteriors on the mass ratio for individual systems to smaller values relative to the true mass ratio.}
\end{figure}

Using the posterior population distribution on the masses $m_1$ and $m_2$, we calculate the corresponding distribution on the mass ratio $q$ in Figure~\ref{fig:PPD}. \added{The dashed blue line in Figure~\ref{fig:PPD} shows the mass ratio distribution in the underlying population, corresponding to the blue $m_1$--$m_2$ distribution of Figure~\ref{fig:m1m2PPD}, while the orange line shows the mass ratio distribution among detected systems, corresponding to the orange $m_1$--$m_2$ distribution of Figure~\ref{fig:m1m2PPD}. While selection effects have a significant effect on the two-dimensional $m_1$--$m_2$ distribution, the 1-dimensional \emph{mass ratio} distribution among detected systems is nearly identical to the mass ratio distribution in the underlying population.}

We expect 90\% of detected events to have their true masses fall within the orange credible region of Figure~\ref{fig:m1m2PPD}. In terms of the mass ratio distribution, we expect that 90\% of detected events will have mass ratios $q > 0.73$, and 99\% of detected events will have mass ratios $q > 0.51$. We can take these predictions one step further by simulating the \emph{measured} mass ratio values for detected events, which accounts for measurement uncertainty in addition to the selection effects. Given the true masses of a detected event, we generate a mock posterior to represent how those masses would be measured in LIGO/Virgo data. The mock posteriors are generated according to the prescription described in Appendix~\ref{sec:mocks}. We summarize the expected mass ratio posteriors from anticipated detections as the green dashed line (median) and shaded band (symmetric 90\% interval) in Figure~\ref{fig:PPD}. We refer to this green band as the ``posterior predictive process." Based on the first ten detections, and assuming that all detections are described by the same population model assumed here, we expect that 90\% of future detections will have recovered mass ratio posteriors that lie within the shaded band. We see that measurement uncertainty plays a significant role in shifting the observed mass ratio posteriors (with the default flat-in-component-mass priors) away from 1 relative to $q_\mathrm{true}$.

\section{Simulations}
\label{sec:simulations}
We expect to have tens more BBH detections by the end of LIGO/Virgo's third observing run in mid-2020, and hundreds of detections within a few more years~\pp{\citep{ObsScen}}. In this section, we explore the expected mass distribution constraints from tens to hundreds of detections under the models considered here. We perform our analysis on mock GW detections that we generate from known underlying distributions. We follow a simplified yet realistic method for generating mock measurements from the underlying population and ensure that the mock primary and secondary masses are measured with uncertainties typical to second-generation GW detectors~\citep{Vitale:2017}. The method for generating mock detections is described in Appendix~\ref{sec:mocks}.

\begin{figure*}
\includegraphics[width=\textwidth]{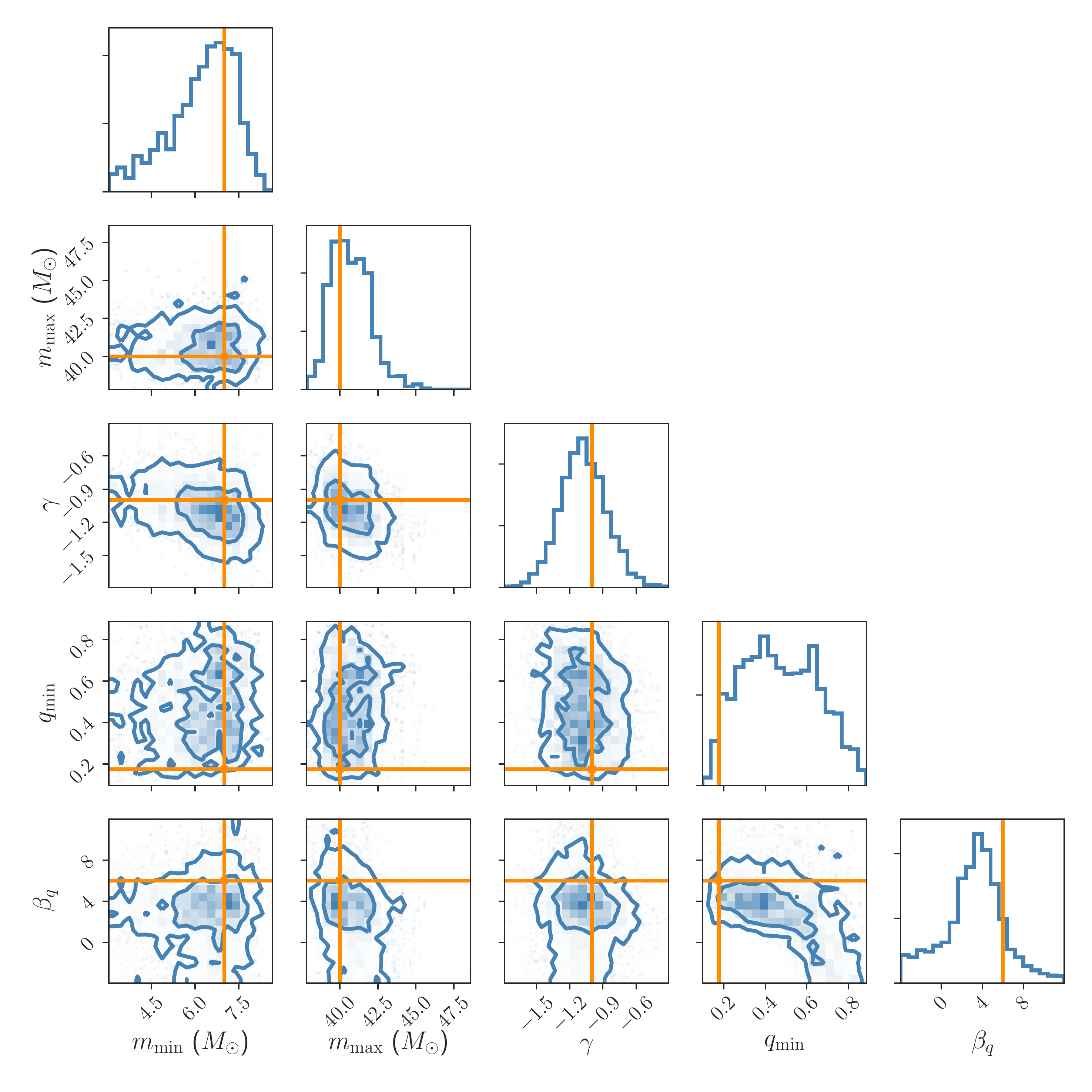}
\caption{\label{fig:mockcorner} Constraints on the population hyper-parameters for a simulated population of 60 BBH detections that follow Eq.~\ref{eq:betaq} with $\mmin = 7 \ M_\odot$, $\mmax = 40 \ M_\odot$, $\gamma = -1$, $\beta_q = 6$, and $q_\mathrm{min} = \mmin/\mmax = 0.175$. These injected hyper-parameter values are shown in orange lines. In the two-dimensional plots, the contours show 50\% and 90\% posterior credible regions.}
\end{figure*}

The expected constraints from 60 detections \pp{\citep[similar to what we expect by the end of O3;][]{ObsScen}} are shown in Figure~\ref{fig:mockcorner} for a simulated population described by Eq.~\ref{eq:betaq} with $\mmin = 7 \ M_\odot$, $\mmax = 40 \ M_\odot$, $\gamma = -1$, $\beta_q = 6$, and $q_\mathrm{min} = \mmin/\mmax \pp{=0.175}$. 
We find with 60 events from this simulated population, we can typically rule out random-pairing with Bayes factors $\gtrsim 1000$. These projections are conservative because the deviations from random-pairing in the chosen mock population are not very large compared to the values of $q_\mathrm{min}$ and $\beta_q$ that are favored by the first ten events.
The parameters that govern the pairing mechanism will become increasingly well-constrained, although with large correlations between them. The simulated 60 events shown here yield $\beta_q = \mocksixtybeta$ and $q_\mathrm{scale} = \mocksixtyyscale$ ($q_\mathrm{min} = \mocksixtyqmin$). Meanwhile, the parameters of the underlying power-law mass distribution will also become well-constrained. With 50 more events, we expect to constrain $\mmax$ to a couple of solar masses (this particular realization yields $\mmax = \mocksixtyMMax$) and the power-law slope $\gamma$ to a 90\% credible interval of $< 1$ ($\gamma = \mocksixtygamma$). Note that if $\beta_M$ is left free in the pairing function, the measurement of $\gamma$ will become less informative, as we constrain a linear combination of $\beta_M$ and $\gamma$ (see Section~\ref{sec:results}).  With 100 events, these constraints will improve roughly as $1/N$ and $1/\sqrt{N}$ for $\mmax$ and $\gamma$, respectively\footnote{Because $\mmax$ is a sharp feature, its measurement converges faster than the typical $1/\sqrt{N}$~\citep[see e.g.][]{2003Natur.424...42C,johnson_2007}. This makes it particularly useful as a feature to constrain cosmology~\citep{2019ApJ...883L..42F}.}: 100 simulated events gives $\mmax = \mockhundredMMax$ and $\gamma = \mockhundredgamma$. It may take more events for the constraints on $\mmin$ to become interesting, because the detector sensitivity is a steep function of BH mass, and most detections are at the high end of the mass function~\citep{BigBlackHoles}, or beyond~\citep{2019arXiv191105882F}. For a flat prior starting at $3\,M_\odot$, 60 mock events give $\mmin = \mocksixtyMMin$. However, the constraints on $\mmin$, like the constraints on $\gamma$, are less informative if $\beta_M$ is allowed to vary.

\section{Conclusion}
\label{sec:conclusion}
We have fit the mass distribution of merging BBHs with a simple model that parameterizes the pairing function between the two components in a binary.
We highlight the importance of comparing the full two-dimensional mass distribution of BBHs, because it is impossible to disentangle the overall BH mass distribution from the pairing function when considering only one-dimensional distributions of the primary/ secondary mass or the mass ratio.

Based on the first ten LIGO/Virgo BBH detections, we conclude that component BHs are not randomly paired in a binary; rather, the pairing likely favors components of comparable masses. We find that it is five times more likely that mergers only take place between equal (to within 5\%) mass BHs than that component BHs are randomly drawn from the same underlying distribution.
Our fits imply that 99\% of mass ratios among the population of merging BBHs are greater (closer to equal mass) than $q_\mathrm{1\%} = \qminbetaqonep$. This is to be compared with an expected value of $q_\mathrm{1 \%} = \RPqonep$ for the random pairing scenario. We predict that among detected BBHs, 90\% will have mass ratios $q > 0.73$, and 99\% will have mass ratios $q>0.51$. Meanwhile, we find no evidence that the pairing function depends on the total mass of the system, contrary to the predictions of some dynamical and primordial BH formation channels~\citep{2006ApJ...637..937O,2018ApJ...854...41K,2019arXiv190103345P}.

\added{The current constraints on the pairing function remain compatible with a range of formation channels, with the exception of those that favor random pairing or a preference for unequal mass ratios~\citep{2019arXiv190201864M}. All binaries detected so far are consistent with equal mass components, which is compatible with predictions from the massive overcontact binary/chemically homogeneous formation channel, in which mass transfer may lead to very nearly equal mass components with minimum mass ratios $q_\mathrm{min} \sim 0.9$~\citep{2016A&A...588A..50M}. However, the current constraints on the pairing function are also compatible with scenarios that more mildly prefer equal mass components, such as common envelope binary evolution, which tends to result in mergers with $q \gtrsim 0.5$~\citep{2018arXiv180605820M}, or dynamical interactions in globular clusters, which tend to result in mergers with median mass ratios $q_{50\%} = 0.9$~\citep{2019ApJ...871...91Z}, consistent with our measurement $q_{50\%} = 0.91^{+0.05}_{-0.17}$. On the other hand, some dynamical channels predict that the pairing function should scale with the total mass of the system. While we cannot rule this out with ten detections, it is 6 times more likely that the pairing function has some mass-ratio dependence rather than depending on total mass alone. 

Although the data does not call for a total-mass dependence, it remains possible that the pairing probability depends on the total mass in addition to the mass ratio. If the merger probability scales with total mass as $M_\mathrm{tot}^{\beta_M}$, the implied power-law slope of the mass distribution among single BHs is roughly $\gamma \sim -\frac{1+\beta_M}{2}$. Thus, prior belief that the BH mass spectrum is steep (with $\gamma \lesssim -2.2$) would suggest that the pairing function depends on the total mass with $\beta_M > 0$.} 

By the end of O3, the details of the pairing function will be better constrained (compare the joint posterior on $q_\mathrm{min}$ and $\beta_q$ in Figure~\ref{fig:corner_qminbeta}---the current constraints---with Figure~\ref{fig:mockcorner}---the constraints we expect by the end of O3). We hope that these results will enable detailed comparisons with the predictions of the full 2-dimensional merger rate $\mathcal{R}(m_1, m_2)$ from population synthesis simulations.

As usual, our results rely on the assumption that there is a single population of BBHs that is adequately described by our simple parameterized model~\citep[see e.g.][]{2019arXiv191104424D}. One way to test the validity of this assumption with future detections is to compare them against the posterior predictive distribution (for example, Figures~\ref{fig:m1m2PPD} and~\ref{fig:PPD}) inferred from the model. We conclude that the universe does not assemble its black-hole binaries at random, and future constraints of the pairing function we have introduced above will yield important insights into these formation processes.

\acknowledgments
We are grateful to the LIGO and Virgo collaborations for publicly releasing posterior samples on the Gravitational Wave Open Science Center. We thank Eve Chase and Reed Essick for their helpful comments on the manuscript, and Reed Essick for suggesting the concept of a posterior predictive process for Figure 4. MF was supported by the NSF Graduate Research Fellowship Program under grant DGE-1746045. MF and DEH were supported by  NSF grant PHY-1708081. They were also supported by the Kavli Institute for Cosmological Physics at the University of Chicago through NSF grant PHY-1125897 and an endowment from the Kavli Foundation. DEH also gratefully acknowledges support from the Marion and Stuart Rice Award.

\appendix
\section{Methods}
\label{sec:methods}
We carry out a hierarchical Bayesian analysis to fit the hyper-parameters for each of the mass models discussed in Section~\ref{sec:models}.
We fit only for the distribution of primary and secondary masses, and fix the distributions of all other BBH intrinsic and extrinsic source parameters. We fix the underlying redshift distribution to follow a merger rate that is uniform in comoving volume and source-frame time. We assume that the underlying population is isotropic on the sky, with isotropic inclination angles. For definiteness we fix the spin distribution of both binary components to be uniform in spin magnitude and isotropic in spin tilt. Although this distribution is not necessarily favored by the data, the correlation between the inferred spin distribution and the inferred mass distribution is negligible, as shown in~\citet{Abbott:pop}, which fit simultaneously for the mass and spin distribution. In particular, despite using a different spin model, we recover the results of~\citet{Abbott:pop} under the same mass model.

The likelihood is given by the inhomogeneous Poisson process likelihood~\citep{Loredo:2004,Mandel:2016,Abbott:pop}.
For $N_\mathrm{obs}$ independent events, the likelihood of the data $d$ given hyper-parameters $\theta$ is:
\begin{equation}
\label{eq:likelihood}
    p(d \mid \theta) \propto e^{-\mu(\theta)}
  \prod_{i=1}^{N_{\mathrm{obs}}}
    \int
      p(d_i | m_1, m_2) \, \frac{\mathrm{d}\mathcal{R}}{\mathrm{d}m_1 m_2}\left(\theta \right) \,
    \mathrm{d} m_1 m_2,
\end{equation}
where $p(d_i \mid m_1, m_2)$ denotes the likelihood of an individual event's data given its component masses, $\frac{\mathrm{d}\mathcal{R}}{\mathrm{d}m_1 m_2}\left(\theta \right)$ is the differential merger rate density, which integrates to the total merger rate density $\mathcal{R}$ and is given by $\mathcal{R}p(m_1,m_2 \mid \theta)$, and $\mu(\theta) = \mathcal{R} \langle VT \rangle_\theta$ denotes the expected number of detections given $\mathcal{R}$ and the sensitive spacetime volume $\langle VT \rangle_\theta$ of the detector network to a given population of BBHs with hyper-parameters $\theta$. 

We assume that the merger rate density is uniform in comoving volume and source-frame time, and calculate $\langle VT \rangle_\theta$ according to a semi-analytic prescription~\citep{1993PhRvD..47.2198F,1996PhRvD..53.2878F}. \pp{Following~\cite{Abbott:pop}}, we assume that a single-detector signal-to-noise (SNR) threshold of 8 is necessary and sufficient for detection, and that the detector's noise curve is described by the Early High Sensitivity power spectral density (PSD) for advanced LIGO~\citep{ObsScen}.
The validity of these assumptions is discussed in~\cite{Abbott:pop}. \pp{Unlike in~\cite{Abbott:pop}}, in this work we do not calibrate the $VT(m_1,m_2)$ to the results of injection campaigns into the detection pipelines. As we demonstrate by explicitly comparing our results to those of~\cite{Abbott:pop} in Section~\ref{sec:results}, using the uncalibrated $VT$ calculation leads to a slight bias in our inference of the overall-merger rate, with the  median shifting by a factor of $\sim 1.7$, as expected from Fig.~9 in \citet{Abbott:pop}. However, this does not affect the inferred shape of the mass distribution, which is our primary interest in this work. We also neglect the effect of non-zero spins in the estimation of $VT$, as spins have a sub-dominant effect on the sensitivity~\citep{Abbott:pop}, especially given that existing detections disfavor a significant population of highly spinning systems.

Note that if we marginalize over the rate density $\mathcal{R}$ with a flat-in-log prior, the likelihood takes the form~\citep{Redshift,Mandel:2016}:
\begin{equation}
    p(d \mid \theta) \propto \prod_{i=1}^{N_\mathrm{obs}} \frac{\int p(d_i \mid m_1, m_2)p(m_1,m_2 \mid \theta) \, \mathrm{d}m_1\mathrm{d}m_2}{\int VT(m_1,m_2)p(m_1,m_2 \mid \theta)\, \mathrm{d}m_1\mathrm{d}m_2}.
\end{equation}

To get the individual-event likelihood term $p(d_i \mid m_1, m_2)$ that appears in Eq.~\ref{eq:likelihood}, we use the publicly available {\tt IMRPhenomPv2} posterior samples for the ten BBH detections in O1 and O2~\citep{CatalogData}. There is a negligible difference between the mass posteriors derived with the {\tt IMRPhenomPv2} waveform\pp{~\citep{PhysRevD.93.044006,2016PhRvD..93d4007K}} and the {\tt SEOBNRv3} waveform\pp{~\citep{2014PhRvD..89h4006P}}. The individual-event posteriors were calculated under priors that are flat in \emph{detector}-frame masses, and ``volumetric" in luminosity distance, $d_L$. In terms of source-frame masses and cosmological redshift $z$, the default event-level prior is therefore~\citep{Abbott:pop}:
\begin{equation}
    p(m_1, m_2, z) \propto d_L(z)^2(1+z)^2\left(d_C(z)+\frac{(1+z)d_H}{E(z)}\right),
\end{equation}
where $d_C$ is the comoving distance and $d_H = c/H_0$ is the Hubble distance, and $E(z) = H(z)/H_0$~\citep{Hogg:cosmo}. We divide out by these priors in our analysis to get a term that is proportional to the likelihood rather than the posterior. We fix the cosmological parameters to the best-fit Planck 2015 values~\citep{Planck:2015} throughout for consistency with~\citet{Abbott:pop} and~\citet{Abbott:catalog}. 

We sample from the overall likelihood of Eq.~\ref{eq:likelihood} using PyMC3~\citep{PyMc3}. In all models considered, we choose priors that are flat over $\mmin$, $\mmax$ and the power-law slope $\gamma$ within the ranges $3 \ M_\odot < \mmin < 10 \ M_\odot$, $35 \ M_\odot < \mmax < 100 \ M_\odot$ and $-4 < \gamma < 2$. We take a flat-in-log prior on the rate $p(\mathcal{R}) \propto 1/\mathcal{R}$. Unless they are fixed to some value, we take a flat prior on $\beta_q$ in the range $-4 < \beta_q < 12$ and a flat prior on $\beta_M$ in the range $ 0 < \beta_M < 12 $. Because the prior range of the minimum mass ratio $q_\mathrm{min}$ depends on two other free parameters, $\mmin$ and  $\mmax$, we introduce another parameter $q_\mathrm{scale}$, defined so that:
\begin{equation}
    q_\mathrm{min} = \mmin / \mmax + q_\mathrm{scale}(0.95-\mmin / \mmax),
\end{equation}
and, unless it is fixed, we sample over $q_\mathrm{scale}$ with a flat prior from 0 to 1, so that $\mmin/\mmax < q_\mathrm{min} < 0.95$. We restrict the upper limit of $q_\mathrm{min}$ to slightly below 1 in order to avoid sampling issues, as the mass ratio of any individual GW event is measured with a finite resolution, and this prevents $q_\mathrm{min}$ from being resolved arbitrarily close to $q_\mathrm{min} = 1$.
\pp{When using Model B from~\cite{Abbott:pop}, we use their same priors, with the exception of lower prior boundary for $\mmin$, which we take to be $3 \ M_\odot$ rather than  $5 \ M_\odot$.} 

For those models which contain random-pairing as a subset, we quantify the evidence for the random-pairing hypothesis versus the full model by calculating the Savage-Dickey density ratio (SDDR), which is defined as the ratio of the posterior probability to the prior probability at the given point in parameter space~\citep{SDDR}. \added{We calculate the evidence between the pure mass-ratio dependent pairing function of Eq.~\ref{eq:betaq} and the pure total-mass dependent pairing function of Eq.~\ref{eq:betam} by introducing a more general model that contains both models of interest as nested models. This general model contains a mixture parameter $x$, where $x$ denotes the amplitude of the mass-ratio dependent component and $(1-x)$ denotes the amplitude of the total-mass dependent component. By sampling from this mixture model, the recovered likelihood at $x = 1$ compared to $x = 0$ denotes the evidence in favor of a pure mass-ratio dependent pairing function.} 

\section{Simulated Detections}
\label{sec:mocks}
In generating mock detections, we assume that the underlying population follows a uniform in comoving volume and source-frame time merger rate, with isotropic sky positions and inclinations, and zero spins. The true component masses are drawn from the given population distribution. We note that the assumptions of fixed redshift and spin distributions are unlikely to affect the inference of the pairing function (mass ratios are measured independently of redshift, and excluding spins did not make a difference in the O1 and O2 analysis); however, these distributions can be fit jointly with the mass distribution and marginalized over~\citep{Abbott:pop}. 

Given the true parameters of the binary, we calculate the SNR of the signal in a single detector, assuming that the noise is described by the Mid-High Sensitivity PSD as expected for O3 for the LIGO detectors~\pp{\citep{ObsScen}}. We assume that the binary is then detected if it passes a single-detector SNR threshold of 8. In order to assign measured component masses to each detected binary, we assume that the fractional uncertainty on the source-frame chirp mass follows $\frac{\sigma_\mathcal{M}}{\mathcal{M}} = \frac{8}{\rho}\left(0.01+\left(\frac{0.2z}{1+z}\right)^2 \right)^{1/2}$, where $z$ is the true redshift, while the uncertainty on the symmetric mass ratio $\eta \equiv \frac{m_1 m_2}{(m_1+m_2)^2}$ follows $\sigma_\eta = 0.03\frac{8}{\rho}$, where $\rho$ is the single-detector SNR of the source. Given a true value of $\mathcal{M}$ and $\eta$ for each binary, we randomly draw $\mathcal{M}_\mathrm{obs}$ from a log-normal distribution centered on $\mathcal{M}$ with standard deviation $\sigma_\mathcal{M}$, and $\eta_\mathrm{obs}$ from a normal distribution centered on $\eta$ with standard deviation $\sigma_\eta$. With these values of $\mathcal{M}_\mathrm{obs}$ and $\eta_\mathrm{obs}$ and their assumed known distributions about the true chirp mass $\mathcal{M}$ and symmetric mass ratio $\eta$, we generate mock posterior samples for the component masses $m_1$ and $m_2$ under flat priors using the Monte-Carlo sampler PyStan~\citep{Stan}. These uncertainties are typical of the O2 detections, and result in typical 90\% measurement uncertainties on the source-frame component masses of $\approx 50\%$, with a distribution of uncertainties that matches the one in~\citet{Vitale:2017}.

\bibliographystyle{aasjournal}
\bibliography{references}

\end{document}